\documentclass[journal=nalefd,manuscript=article]{achemso}

\usepackage{graphicx}
\usepackage[svgnames]{xcolor}
\usepackage{mathtools}
\usepackage{hyperref}
\hypersetup{pdftitle={PolyX-H},
pdfauthor={ICN2},
pdfnewwindow=true, 
colorlinks=true,   
linkcolor=DarkBlue,     
citecolor=DarkBlue,   
filecolor=DarkBlue, 
urlcolor=DarkBlue      
}
\usepackage{achemso}

\setkeys{acs}{super=true}
\setkeys{acs}{etalmode=truncate}
\setkeys{acs}{hyperref=true}
\setkeys{acs}{maxauthors=6}

\DeclareMathOperator{\Tr}{Tr}

\author{Jose E. Barrios Vargas}
\email{jose.barrios@icn2.cat}
\affiliation{Catalan Institute of Nanoscience and Nanotechnology (ICN2), CSIC and The Barcelona Institute of Science and Technology, Campus UAB, 08193 Barcelona, Spain}
\altaffiliation{These authors contributed equally to this work.}

\author{Jesper T. Falkenberg}
\affiliation{Department of Micro- and Nanotechnology (DTU Nanotech), Center for Nanostructured Graphene (CNG), Technical University of Denmark, DK-2800 Kgs. Lyngby, Denmark}
\altaffiliation{These authors contributed equally to this work.}

\author{David Soriano}
\affiliation{Catalan Institute of Nanoscience and Nanotechnology (ICN2), CSIC and The Barcelona Institute of Science and Technology, Campus UAB, 08193 Barcelona, Spain}

\author{Aron W. Cummings}
\affiliation{Catalan Institute of Nanoscience and Nanotechnology (ICN2), CSIC and The Barcelona Institute of Science and Technology, Campus UAB, 08193 Barcelona, Spain}

\author{Mads Brandbyge}
\affiliation{Department of Micro- and Nanotechnology (DTU Nanotech), Center for Nanostructured Graphene (CNG), Technical University of Denmark, DK-2800 Kgs. Lyngby, Denmark}

\author{Stephan Roche}
\email{stephan.roche@icn2.cat}
\affiliation{Catalan Institute of Nanoscience and Nanotechnology (ICN2), CSIC and The Barcelona Institute of Science and Technology, Campus UAB, 08193 Barcelona, Spain}
\alsoaffiliation{ICREA - Instituci\'{o} Catalana de Recerca i Estudis Avan\c{c}ats, 08010 Barcelona, Spain}

\title
{Grain boundary-induced variability of charge transport in hydrogenated polycrystalline graphene}

\begin{document}

\begin{abstract}
Chemical functionalization has proven to be a promising means of tailoring the unique properties of graphene. For example, hydrogenation can yield a variety of interesting effects, including a metal-insulator transition or the formation of localized magnetic moments. Meanwhile, graphene grown by chemical vapor deposition is the most suitable for large-scale production, but the resulting material tends to be polycrystalline. Up to now there has been relatively little focus on how chemical functionalization, and hydrogenation in particular, impacts the properties of polycrystalline graphene. In this work, we use numerical simulations to study the electrical properties of hydrogenated polycrystalline graphene. We find a strong correlation between the spatial distribution of the hydrogen adsorbates and the charge transport properties. Charge transport is weakly sensitive to hydrogenation when adsorbates are confined to the grain boundaries, while a uniform distribution of hydrogen degrades the electronic mobility. This difference stems from the formation of the hydrogen-induced resonant impurity states, which are inhibited when the honeycomb symmetry is locally broken by the grain boundaries. These findings suggest a tunability of electrical transport of polycrystalline graphene through selective hydrogen functionalization, and also have implications for hydrogen-induced magnetization and spin lifetime of this material.
\end{abstract}

\section{Introduction}
Since its experimental isolation in 2004~\cite{Novoselov2004}, single-layer graphene has emerged as an exciting material for a wide variety of applications. Much of this excitement stems from graphene's remarkable electrical~\cite{Novoselov2004}, optical~\cite{Nair2008}, thermal~\cite{Baladin2008}, and mechanical properties~\cite{Lee2008}. In addition to its unique intrinsic properties, another promising characteristic of graphene is its tunability. In particular, because graphene is two-dimensional, chemical functionalization has been studied as an effective approach to extrinsically tailor its material properties. For example, metallic adatoms can potentially induce a strong spin-orbit coupling in graphene~\cite{Weeks2011} and oxygen adsorption can significantly alter graphene's thermoelectric characteristics~\cite{Zhang2014b}. An adsorbate of particular interest is hydrogen, which forms a covalent bond to a single carbon atom and induces a resonant impurity state around the graphene Dirac point~\cite{Robinson2008,Wehling2009}. This can have a considerable impact on electronic transport, as revealed by measurements of a metal-insulator transition with increasing hydrogen density~\cite{Ruwantha2013}. Recent experimental work has also shown that localized magnetic moments are formed around hydrogen impurities~\cite{Herrero2016}, which could have important implications for graphene spintronics~\cite{Roche2015,Soriano2015}.

While mechanical exfoliation tends to yield the highest-quality graphene samples in the laboratory, chemical vapor deposition (CVD) is the most efficient method to produce graphene on an industrial scale. This method is now capable of producing single graphene grains reaching the centimeter scale~\cite{Li2015, Lin2016, Wu2015}, but faster CVD growth yields much smaller grains, resulting in a material that is polycrystalline~\cite{Cummings2014,Wu2015}. In polycrystalline graphene, the grain boundaries (GBs) between misoriented grains consist of a series of non-hexagonal rings~\cite{Mesaros2010, Huang2011, Kim2011} that can impede charge transport through the 
material~\cite{Yazyev2010,Yu2011,Tsen2012,Koepke2013,Gargiulo2014nanolett,Nguyen2016}. In addition, GBs tend to be more chemically reactive than pristine graphene, which can also strongly alter charge transport, opening new perspectives for gas sensing applications~\cite{Salehi2012, Yasaei2014}. Prior studies have examined the impact of hydrogenation on the electronic transport properties of polycrystalline graphene~\cite{Ruwantha2013,Cummings2014,Seifert2015}, but the detailed nature of the interaction between GBs and hydrogen adsorbates remains unclear.

In this work, we use {\it ab initio} and tight-binding (TB) calculations to study the impact of hydrogenation on the electronic properties of polycrystalline graphene. We find that the precise distribution of hydrogen adatoms is crucial for predicting their effect. Specifically, when the hydrogenation is confined to the GBs, the overall impact on charge transport is negligible, which is in sharp contrast to the case of hydrogenation within the grains. We find that this difference is related to the formation (or not) of resonant impurity states formed near the Dirac point; hydrogen adsorbates induce resonant states within the graphene grains but not in the GBs. These results suggest the possibility of tuning the electrical transport of polycrystalline graphene through selective hydrogen functionalization, which could have important implications for hydrogen-induced magnetotransport properties.

\section{Hydrogenation of a Stone-Wales defect}
\begin{figure*}
\includegraphics[width=1.0\linewidth]{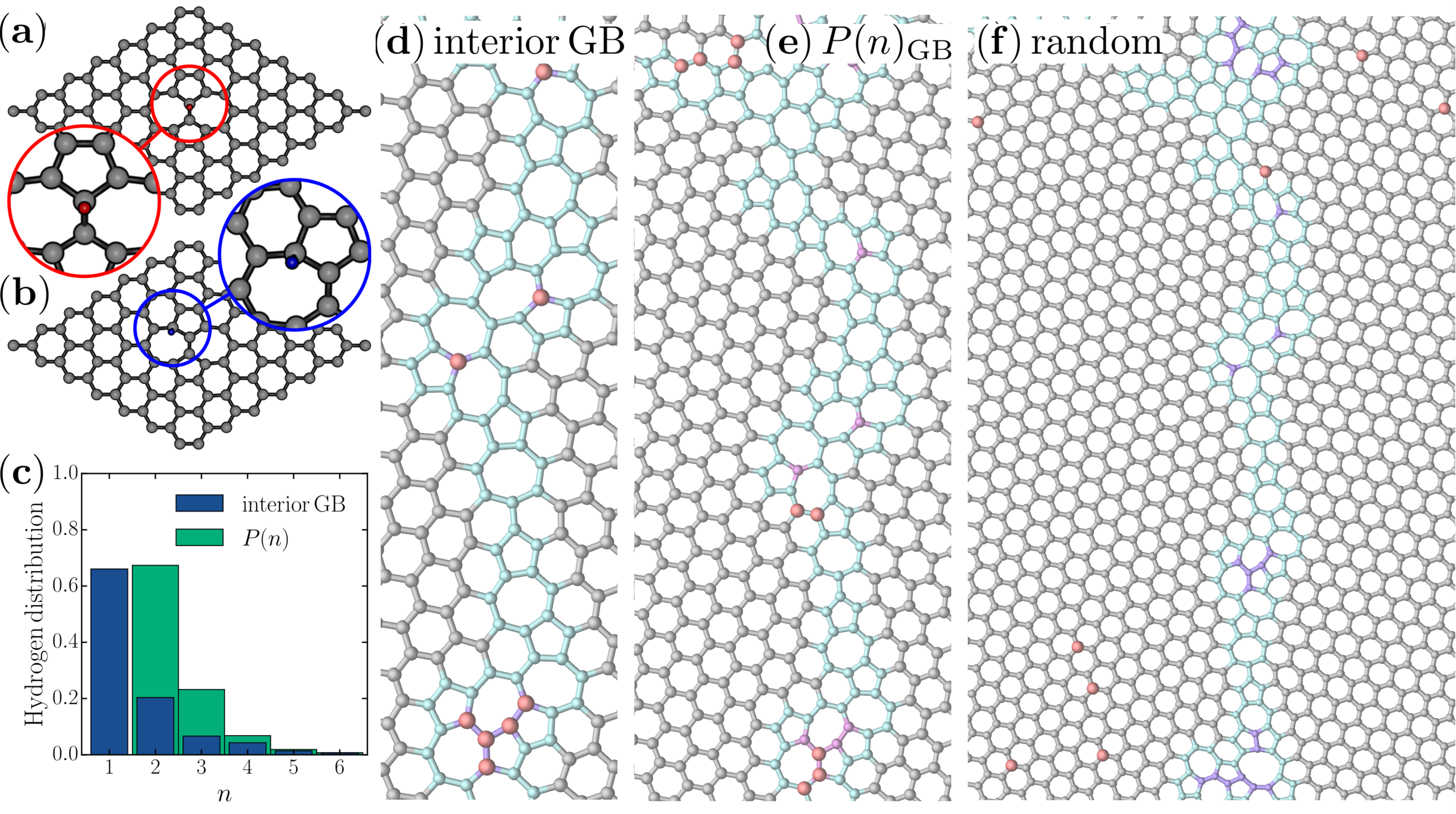}
\caption{Structural configuration of hydrogen adsorption in polycrystalline graphene. Panels (a) and (b) depict H-adsorption on an interior and exterior site of a Stone-Wales defect, respectively. Panel (c) shows the cluster distribution of interior GB sites in the polycrystalline structure and H-segregation following distributions $P(n)$ obtained by Monte Carlo simulations given for pristine graphene (Ref.~\cite{Gargiulo2014}). Panels (d)-(f) show the case of polycrystalline graphene, with adsorption (c) on the interior GB sites, (d) the exterior GB sites, and (e) randomly distributed throughout the sample.}
\label{Samples}
\end{figure*}

We begin our study with a canonical structural defect in graphene, the Stone-Wales (SW) defect. As shown in Figures \ref{Samples}(a) and (b), a SW defect consists of a 90-degree rotation of a single carbon-carbon bond, turning four hexagons into two pairs of pentagons and heptagons. For this defect, and for all GB structures in general, we classify ``interior'' defect sites as the carbon sites that only belong to non-hexagonal rings, and ``exterior'' defect sites as those that belong to both hexagonal and non-hexagonal rings. By this definition, the exterior sites lie on the boundary between the SW defect (or the GB) and the pristine graphene region. Various {\it ab initio} calculations have shown that the interior sites are more favorable for chemical adsorption~\cite{Boukhvalov2008, OuYang2008}. To study the impact of hydrogenation, we calculate the electronic band structure of the SW defect with a single hydrogen impurity at either an interior or an exterior defect site, as shown schematically in Figures \ref{Samples}(a) and (b). 

The electronic structure calculations were performed using the SIESTA {\it ab initio} package~\cite{Soler2002}. We use the Perdew-Burke-Ernzerhof (PBE) exchange-correlation functional within the generalized gradient approximation (GGA) \cite{Perdew1996}. The $7 \times 7$ supercells containing the hydrogenated Stone-Wales defects were fully relaxed using a $8 \times 8$ k-point sampling, and employing a double-$\zeta$ polarized basis set.      

\begin{figure}
\begin{tabular}{c}
\includegraphics[width=1.0\linewidth]{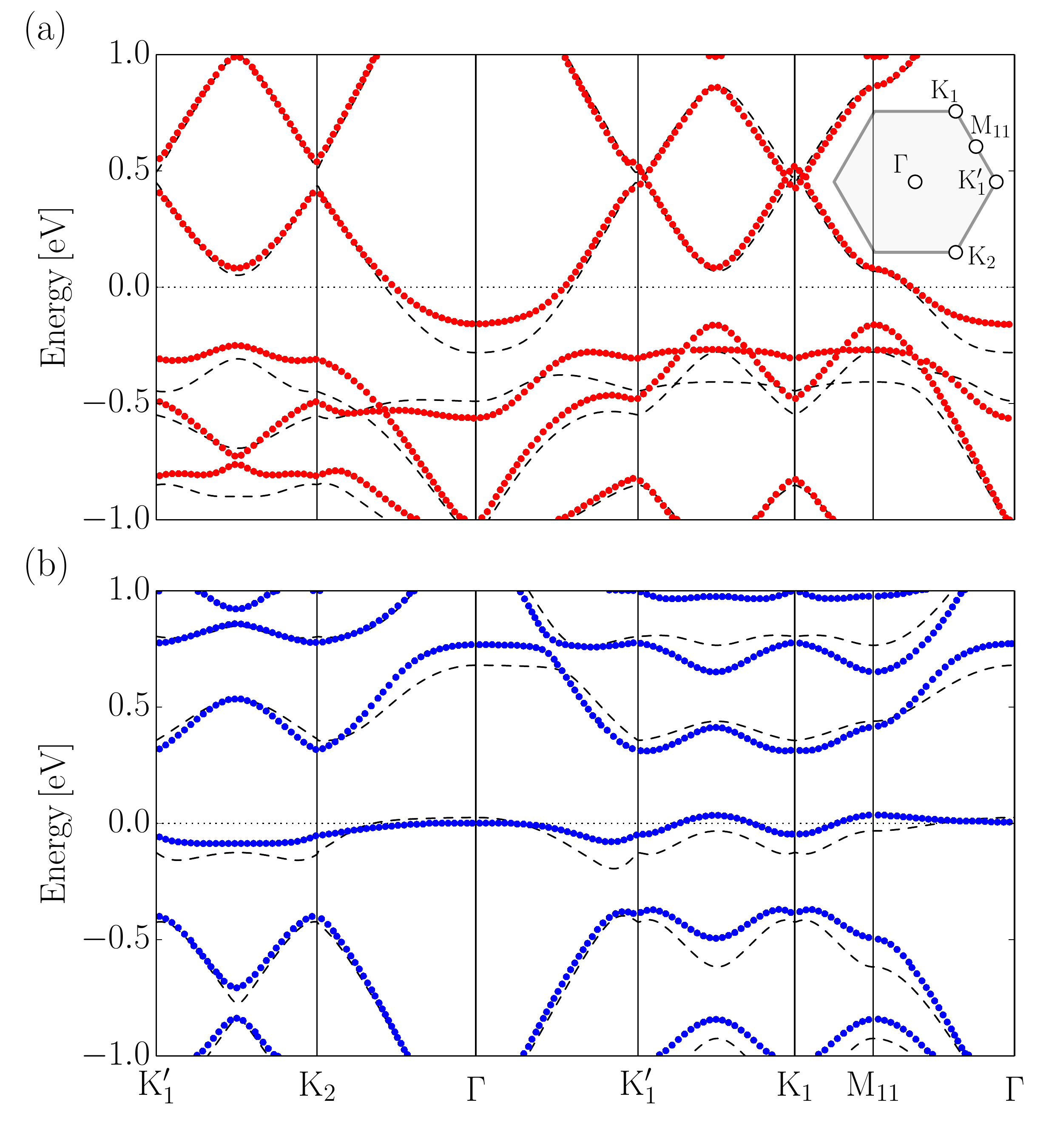}
\end{tabular}
\caption{Bandstructure of a hydrogenated Stone-Wales defect. Panel (a) is for adsorption on an interior site, while panel (b) shows the case of adsorption on an exterior defect site. Solid (dashed) lines are obtained from {\it ab initio} (tight-binding) calculations.}
\label{Bands}
\end{figure}

Resulting band structures, shown in Figures \ref{Bands}(a) and (b), indicate a clear difference between the two cases. As seen in Figure \ref{Bands}(b), the hydrogenation of an exterior site opens a band gap and induces a strongly localized impurity state around the Fermi energy, similar to what is seen in hydrogenated pristine graphene~\cite{Robinson2008, Wehling2009}. In constrast, Figure \ref{Bands}(a) reveals that the localized impurity state is completely suppressed when hydrogenating the interior defect site. 

These results echo those reported by other groups~\cite{Duplock2004,Brito2011}, evidencing the importance of the local atomic structure in determining the formation of localized impurity states. According to graph theory, the number of zero-energy eigenvalues in a bipartite lattice is given by $n_0 = |n_A - n_B|$, where $n_A$ and $n_B$ are the number of sites in each sublattice~\cite{Lieb1989,PhysRevB.49.3190}. At all carbon sites around the SW defect, the bipartite nature of the graphene lattice is preserved and hydrogenation will induce an impurity state at zero energy (the Dirac point). However, the two interior atoms of the SW defect are each connected to both sublattices of the surrounding bipartite lattice, and thus they cannot be assigned to either of the two sublattices. In this case, hydrogenation does not induce an imbalance of the two sublattices, and the zero-energy impurity state does not form.

\section{Hydrogenation of polycrystalline graphene}
Moving beyond the SW defect, we now consider hydrogenation of a more realistic polycrystalline graphene sample. For this work, a large-area polycrystalline sample containing $\sim$2.2 million atoms was generated according to the method of Ref.~\cite{VanTuan2013}, with an average grain diameter of 21 nm. Owing to its size, the electronic properties of this sample were described by a nearest-neighbor tight-binding (TB) model with a single $p_z$-orbital per carbon site. As shown by the dashed lines in Figure~\ref{Bands}, this simple model well reproduces the {\it ab initio} calculations of the SW defect. To calculate electronic transport in the polycrystalline sample, we employed a real-space order-N wave packet propagation method~\cite{Roche1999, RocheBook}. Through this method one can calculate the time-dependent diffusion coefficient as

\begin{equation}
D(E,t) = \frac{\partial}{\partial t}\Delta X^2(E,t),
\label{diffusion}
\end{equation}

\noindent where $\Delta X^2$ is the mean-square displacement of the wave packet,

\begin{equation}
\Delta X^2(E,t) = \frac{\Tr[\delta (E-\hat{H}) |\hat{X}(t) - \hat{X}(0)|^2]}{\rho(E)},
\label{msd}
\end{equation}

\noindent and $\rho(E) = \Tr[\delta (E-\hat{H})]$ is the density of states (DOS), which is computed using a random phase vector~\cite{Iitaka2004} and 
the kernel polynomial method~\cite{Weisse2006}. Finally, the semiclassical conductivity, mean free path, and mobility were calculated as $\sigma(E) = e^2 \rho(E) D_{max}(E)$, $\ell_e(E) = D_{max}(E) / 2v_F(E)$, and $\mu(E) = \sigma(E) / n(E)$, where $D_{max}$ is the maximum value of the time-dependent diffusion coefficient, $v_F$ is the electron Fermi velocity, and $n$ is the charge density, obtained by integrating the DOS.

\begin{figure}
\begin{tabular}{c}
\includegraphics[width=0.8\linewidth]{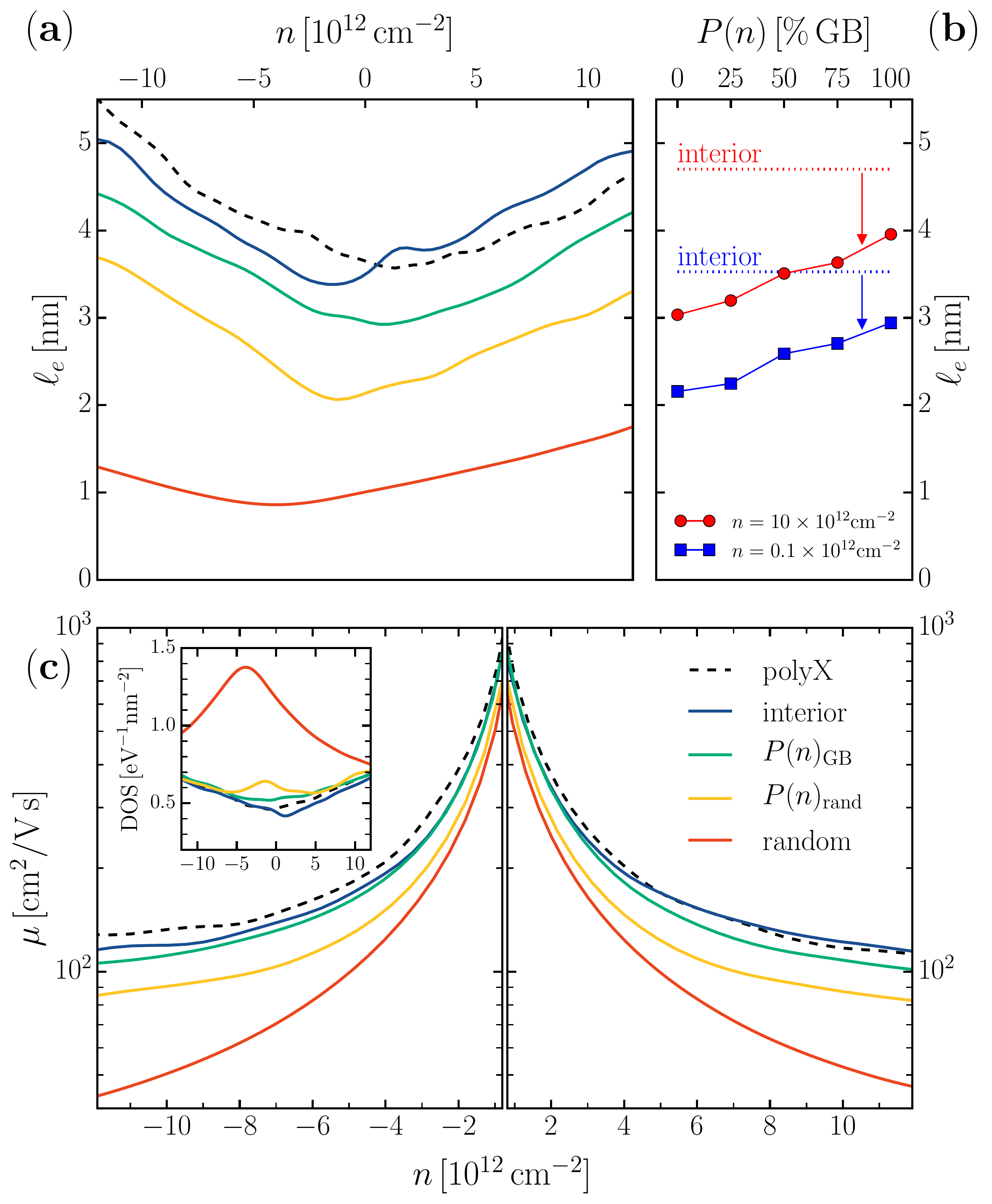}
\end{tabular}
\caption{Charge mobility in hydrogenated polycrystalline graphene. Mean free path versus carrier density for four different hydrogen distributions (panel (a)), and for different weights of hydrogen in the grains and the GBs, assuming a global distribution of $P(n)$ (panel (b)).  Panel (c) shows the mobility (main frame) and the related DOS (inset) for the four hydrogen distributions. In panels (a) and (c), the blue, green, yellow, and red lines are for hydrogenation of the interior GB sites, $P(n)$ in the GB sites, $P(n)$ throughout the sample, and a random distribution of hydrogen throughout the sample, respectively, while the dashed line is for the hydrogen-free polycrystalline sample. In all cases, the hydrogen density is 0.35\%, corresponding to 100\% occupation of the interior GB sites.}
\label{MFPdist}
\end{figure}

Monte Carlo simulations have shown that at room temperature, hydrogen adatoms tend to aggregate in clean graphene with a cluster distribution, $P(n)$ manifesting a peak at hydrogen dimers~\cite{Gargiulo2014}. However, the binding energies of hydrogen at GBs are significantly higher than in pristine graphene~\cite{Brito2011,Zhang2014a}. Thus, we expect that in polycrystalline graphene the GBs remain the most likely site for hydrogen adsorption, but with some higher fraction of dimers forming inside the grains. To highlight the importance of such variability in hydrogen distribution, we consider four different functionalization situations in our simulations, with hydrogens 1) restricted to the interior GB sites, 2) distributed among all the GB sites according to $P(n)$, 3) distributed throughout the sample according to $P(n)$, and 4) distributed completely randomly throughout the sample. These distributions are shown schematically in Figures~\ref{Samples}(d)-(f), where we zoom in on one particular GB. In Fig.~\ref{Samples}(c), we also show the cluster distribution $P(n)$ and compare it to the cluster distribution of the interior GB sites. Here we see that most of the interior GB sites are isolated.

Figure~\ref{MFPdist} shows the results of our transport calculations assuming a hydrogen density of 0.35\%, which is the amount needed to fully saturate the interior GB sites. The blue, green, yellow and red lines are for hydrogenation of the interior GB sites, $P(n)$ in the GB sites, $P(n)$ throughout the sample, and a random distribution throughout the sample, respectively. The dashed line is for the polycrystalline sample in the absence of hydrogen. In Figure~\ref{MFPdist}(a), there is a clear correlation between the hydrogen distribution and the resulting decrease in mean free path, $\ell_e$, which can be also seen as an increase in the DOS around the charge neutrality point (see inset Figure~\ref{MFPdist}(c)). In particular, for a uniform distribution of hydrogen, a resonant peak in the DOS appears near the Dirac point, which is the signature of hydrogenation of pristine graphene. However, when the hydrogenation is confined to the GBs, this peak is strongly suppressed for the $P(n)$ distribution and is completely absent for the interior sites. Actually, for interior site hydrogenation, the DOS actually {\it decreases} on the electron side compared to the polycrystalline sample without hydrogen (polyX). Here,  an uniform hydrogenation significantly reduces $\ell_e$ and $\mu$ (Figure~\ref{MFPdist}(c)), while hydrogenation following the $P(n)$ distribution has a smaller impact. Meanwhile, hydrogenation of the interior GB sites appears to have, on average, little to no impact on the electrical transport properties of polycrystalline graphene. Note that for the same hydrogen density, the mobility roughly differs by a factor of 3 between the homogeneous and inhomogeneous adsorbate distributions. By gradually varying the spatial distribution of dimers, trimers, tetramers, etc. from the grains to the GB sites, we observe that $\ell_e$ increases linearly for a fixed charge carrier concentration but differs substantially from the interior GB case (see Figure~\ref{MFPdist}(b)).

\begin{figure}
\includegraphics[width=1.0\linewidth]{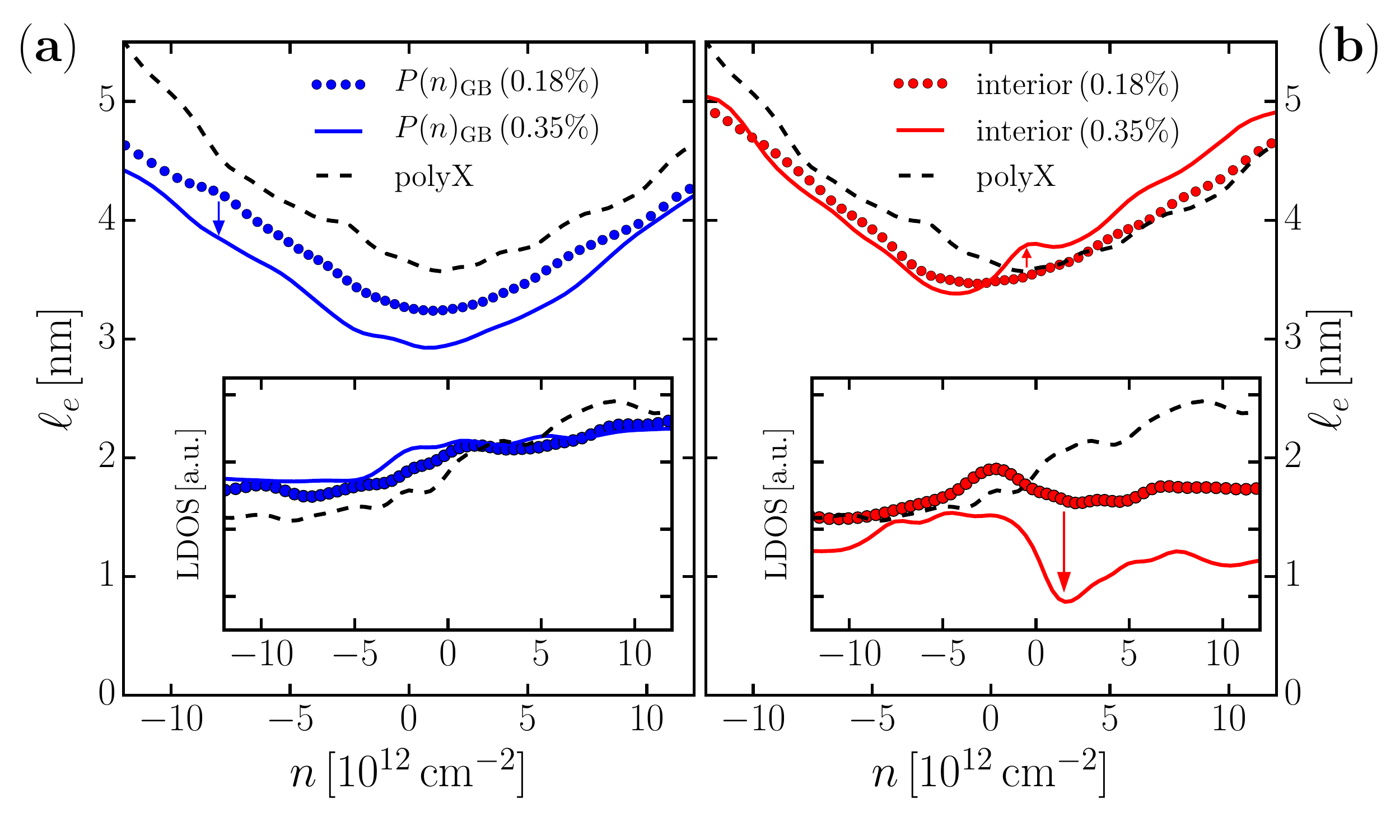}
\caption{Hydrogenation of grain boundary sites. Panel (a) shows the mean free path for hydrogenation using the $P(n)$ distribution on all GB sites, while panel (b) is for hydrogenation of the interior GB sites. The solid line is for a hydrogen density of 0.35\%, corresponding to full saturation of the interior grain boundary sites. The dotted line is for a hydrogen density of 0.18\%, and the dashed curve is for the polycrystalline sample without hydrogenation. The insets show the local density of states, projected over the grain boundary atoms.}
\label{MFPconc}
\end{figure}

Figure \ref{MFPconc} shows the impact of varying the hydrogen density on the graphene GBs. The solid line is for a hydrogen density of 0.35\%, the dotted line is for 0.18\%, and the dashed line is for the absence of hydrogen. The main panels of Figure~\ref{MFPconc} reveal the trends suggested in Figure~\ref{MFPdist}; for a $P(n)$ distribution on GB sites the mean free path is reduced with an increasing density of hydrogen (Figure~\ref{MFPconc}(a)), while hydrogenation of the interior GB sites has, on average, a negligible impact (Figure~\ref{MFPconc}(b)). However, as noted before, there is an unexpected {\it increase} of the mean free path on the electron side when hydrogenating the interior GB sites. This behavior is also reflected in the local DOS (LDOS) of the GB atoms, shown in the insets. Here one can see that increasing the hydrogenation of the interior GB sites actually decreases the LDOS of the GBs at certain energies. Thus, it appears that hydrogen adsorbates can passivate the defect states associated with graphene GBs, slightly reducing their adverse impact on charge transport. This only appears to be true for the interior GB sites, as the LDOS always increases (and $\ell_e$ decreases) with increasing hydrogenation of the exterior GB sites.

\section{Discussion and conclusions}
To summarize, we have used {\it ab initio} and tight-binding calculations to study charge transport in hydrogenated polycrystalline graphene. Our calculations reveal that H-impurity states are suppressed when defects are adsorbed on grain boundary sites, thereby modulating the global charge transport features of polycrystalline graphene. Earlier work that studied the impact of clusterization of hydrogen on pristine graphene found a similar result -- clusterization tends to suppress the formation of the zero-energy states which improves the electrical conduction capability~\cite{Gargiulo2014}. It is possible to question the clusterization of hydrogen due to the fact that hydrogen desorbs quickly at room temperature in pristine graphene; but this discussion is beyond the scope of the work. The nature of hydrogen distribution has also important implications for experimental characterization of grain boundaries~\cite{Seifert2015}. Furthermore, theoretically there are many other adsorbates besides hydrogen that can give rise to resonant states, including PMMA, which is typically used to transfer graphene to a substrate~\cite{Santos2012}. Given its relatively large size, it seems unlikely that PMMA would form dimers.\\

Our results can also have important consequences for graphene spintronics \cite{Roche2015}. Recent experiment has shown that the resonant state induced by a hydrogen adsorbate is magnetic, with an exchange splitting of $\sim$20 meV for graphene grown on SiC~\cite{Herrero2016}. This local, magnetic resonant state can strongly alter spin relaxation times in graphene~\cite{Kochan2014,Soriano2015}, and is undesirable for the development of graphene-based spintronic devices. The suppression of the resonant state at the GBs suggests that spin lifetimes in polycrystalline graphene may be unaffected by a small density of hydrogen impurities. Finally, the presence of grain boundaries could also modulate the magnetoresistance signals predicted for paramagnetic, antiferromagnetic, or ferromagnetic macroscopic states~\cite{Soriano2011}.

\begin{acknowledgement}
This work has received funding from the European Union Seventh Framework Programme under grant agreement 604391 Graphene Flagship. S.R. acknowledges Funding from the Spanish Ministry of Economy and Competitiveness and the European Regional Development Fund (Project No. FIS2015-67767-P (MINECO/FEDER)), the Secretaria de Universidades e Investigacion del Departamento de Economia y Conocimiento de la Generalidad de Cataluna and the Severo Ochoa Program (MINECO SEV-2013-0295). The Center for Nanostructured Graphene (CNG) is sponsored by the Danish National Research Foundation, Project DNRF103. J.E.B.-V. acknowledges support from CONACyT (Mexico, D.F.). JTF acknowledge support from the Lundbeck foundation (R95-A10510).
\end{acknowledgement}

\bibliography{BibGBsites}{}

\providecommand{\latin}[1]{#1}
\providecommand*\mcitethebibliography{\thebibliography}
\csname @ifundefined\endcsname{endmcitethebibliography}
  {\let\endmcitethebibliography\endthebibliography}{}
\begin{mcitethebibliography}{47}
\providecommand*\natexlab[1]{#1}
\providecommand*\mciteSetBstSublistMode[1]{}
\providecommand*\mciteSetBstMaxWidthForm[2]{}
\providecommand*\mciteBstWouldAddEndPuncttrue
  {\def\EndOfBibitem{\unskip.}}
\providecommand*\mciteBstWouldAddEndPunctfalse
  {\let\EndOfBibitem\relax}
\providecommand*\mciteSetBstMidEndSepPunct[3]{}
\providecommand*\mciteSetBstSublistLabelBeginEnd[3]{}
\providecommand*\EndOfBibitem{}
\mciteSetBstSublistMode{f}
\mciteSetBstMaxWidthForm{subitem}{(\alph{mcitesubitemcount})}
\mciteSetBstSublistLabelBeginEnd
  {\mcitemaxwidthsubitemform\space}
  {\relax}
  {\relax}

\bibitem[Novoselov \latin{et~al.}(2004)Novoselov, Geim, Morozov, Jiang, Zhang,
  Dubonos, Grigorieva, and Firsov]{Novoselov2004}
Novoselov,~K.~S.; Geim,~A.~K.; Morozov,~S.~V.; Jiang,~D.; Zhang,~Y.;
  Dubonos,~S.~V. \latin{et~al.}  \emph{Science} \textbf{2004}, \emph{306},
  666--669\relax
\mciteBstWouldAddEndPuncttrue
\mciteSetBstMidEndSepPunct{\mcitedefaultmidpunct}
{\mcitedefaultendpunct}{\mcitedefaultseppunct}\relax
\EndOfBibitem
\bibitem[Nair \latin{et~al.}(2008)Nair, Blake, Grigorenko, Novoselov, Booth,
  Stauber, Peres, and Geim]{Nair2008}
Nair,~R.~R.; Blake,~P.; Grigorenko,~A.~N.; Novoselov,~K.~S.; Booth,~T.~J.;
  Stauber,~T. \latin{et~al.}  \emph{Science} \textbf{2008}, \emph{320},
  1308--1308\relax
\mciteBstWouldAddEndPuncttrue
\mciteSetBstMidEndSepPunct{\mcitedefaultmidpunct}
{\mcitedefaultendpunct}{\mcitedefaultseppunct}\relax
\EndOfBibitem
\bibitem[Balandin \latin{et~al.}(2008)Balandin, Ghosh, Bao, Calizo,
  Teweldebrhan, Miao, and Lau]{Baladin2008}
Balandin,~A.~A.; Ghosh,~S.; Bao,~W.; Calizo,~I.; Teweldebrhan,~D.; Miao,~F.
  \latin{et~al.}  \emph{Nano Lett.} \textbf{2008}, \emph{8}, 902--907\relax
\mciteBstWouldAddEndPuncttrue
\mciteSetBstMidEndSepPunct{\mcitedefaultmidpunct}
{\mcitedefaultendpunct}{\mcitedefaultseppunct}\relax
\EndOfBibitem
\bibitem[Lee \latin{et~al.}(2008)Lee, Wei, Kysar, and Hone]{Lee2008}
Lee,~C.; Wei,~X.; Kysar,~J.~W.; Hone,~J. \emph{Science} \textbf{2008},
  \emph{321}, 385--388\relax
\mciteBstWouldAddEndPuncttrue
\mciteSetBstMidEndSepPunct{\mcitedefaultmidpunct}
{\mcitedefaultendpunct}{\mcitedefaultseppunct}\relax
\EndOfBibitem
\bibitem[Weeks \latin{et~al.}(2011)Weeks, Hu, Alicea, Franz, and Wu]{Weeks2011}
Weeks,~C.; Hu,~J.; Alicea,~J.; Franz,~M.; Wu,~R. \emph{Phys. Rev. X}
  \textbf{2011}, \emph{1}, 021001\relax
\mciteBstWouldAddEndPuncttrue
\mciteSetBstMidEndSepPunct{\mcitedefaultmidpunct}
{\mcitedefaultendpunct}{\mcitedefaultseppunct}\relax
\EndOfBibitem
\bibitem[Zhang \latin{et~al.}(2014)Zhang, Fonseca, and Cho]{Zhang2014b}
Zhang,~H.; Fonseca,~A.~F.; Cho,~K. \emph{J. Phys. Chem. C} \textbf{2014},
  \emph{118}, 1436--1442\relax
\mciteBstWouldAddEndPuncttrue
\mciteSetBstMidEndSepPunct{\mcitedefaultmidpunct}
{\mcitedefaultendpunct}{\mcitedefaultseppunct}\relax
\EndOfBibitem
\bibitem[Robinson \latin{et~al.}(2008)Robinson, Schomerus, Oroszl\'{a}ny, and
  Fal'ko]{Robinson2008}
Robinson,~J.~P.; Schomerus,~H.; Oroszl\'{a}ny,~L.; Fal'ko,~V.~I. \emph{Phys.
  Rev. Lett.} \textbf{2008}, \emph{101}, 196803\relax
\mciteBstWouldAddEndPuncttrue
\mciteSetBstMidEndSepPunct{\mcitedefaultmidpunct}
{\mcitedefaultendpunct}{\mcitedefaultseppunct}\relax
\EndOfBibitem
\bibitem[Wehling \latin{et~al.}(2009)Wehling, Katsnelson, and
  Lichtenstein]{Wehling2009}
Wehling,~T.~O.; Katsnelson,~M.~I.; Lichtenstein,~A.~I. \emph{Phys. Rev. B}
  \textbf{2009}, \emph{80}, 085428\relax
\mciteBstWouldAddEndPuncttrue
\mciteSetBstMidEndSepPunct{\mcitedefaultmidpunct}
{\mcitedefaultendpunct}{\mcitedefaultseppunct}\relax
\EndOfBibitem
\bibitem[Jayasingha \latin{et~al.}(2013)Jayasingha, Sherehiy, Wu, and
  Sumanasekera]{Ruwantha2013}
Jayasingha,~R.; Sherehiy,~A.; Wu,~S.-Y.; Sumanasekera,~G.~U. \emph{Nano Lett.}
  \textbf{2013}, \emph{13}, 5098--5105\relax
\mciteBstWouldAddEndPuncttrue
\mciteSetBstMidEndSepPunct{\mcitedefaultmidpunct}
{\mcitedefaultendpunct}{\mcitedefaultseppunct}\relax
\EndOfBibitem
\bibitem[Gonz\'{a}lez-Herrero \latin{et~al.}(2016)Gonz\'{a}lez-Herrero,
  G\'{o}mez-Rodr\'{i}guez, Mallet, Moaied, Palacios, Salgado, Ugeda, Veuillen,
  Yndurain, and Brihuega]{Herrero2016}
Gonz\'{a}lez-Herrero,~H.; G\'{o}mez-Rodr\'{i}guez,~J.~M.; Mallet,~P.;
  Moaied,~M.; Palacios,~J.~J.; Salgado,~C. \latin{et~al.}  \emph{Science}
  \textbf{2016}, \emph{352}, 437--441\relax
\mciteBstWouldAddEndPuncttrue
\mciteSetBstMidEndSepPunct{\mcitedefaultmidpunct}
{\mcitedefaultendpunct}{\mcitedefaultseppunct}\relax
\EndOfBibitem
\bibitem[Roche \latin{et~al.}(2015)Roche, Akerman, Beschoten, Charlier,
  Chshiev, Dash, Dlubak, Fabian, Fert, Guimarães, Guinea, Grigorieva,
  Schönenberger, Seneor, Stampfer, Valenzuela, Waintal, and van
  Wees]{Roche2015}
Roche,~S.; Akerman,~J.; Beschoten,~B.; Charlier,~J.-C.; Chshiev,~M.;
  Dash,~S.~P. \latin{et~al.}  \emph{2D Materials} \textbf{2015}, \emph{2},
  030202\relax
\mciteBstWouldAddEndPuncttrue
\mciteSetBstMidEndSepPunct{\mcitedefaultmidpunct}
{\mcitedefaultendpunct}{\mcitedefaultseppunct}\relax
\EndOfBibitem
\bibitem[Soriano \latin{et~al.}(2015)Soriano, Tuan, Dubois, Gmitra, Cummings,
  Kochan, Ortmann, Charlier, Fabian, and Roche]{Soriano2015}
Soriano,~D.; Tuan,~D.~V.; Dubois,~S. M.-M.; Gmitra,~M.; Cummings,~A.~W.;
  Kochan,~D. \latin{et~al.}  \emph{2D Materials} \textbf{2015}, \emph{2},
  022002\relax
\mciteBstWouldAddEndPuncttrue
\mciteSetBstMidEndSepPunct{\mcitedefaultmidpunct}
{\mcitedefaultendpunct}{\mcitedefaultseppunct}\relax
\EndOfBibitem
\bibitem[Li \latin{et~al.}(2015)Li, Wang, Liu, Jin, Wang, and Wan]{Li2015}
Li,~J.; Wang,~X.-Y.; Liu,~X.-R.; Jin,~Z.; Wang,~D.; Wan,~L.-J. \emph{J. Mater.
  Chem. C} \textbf{2015}, \emph{3}, 3530--3535\relax
\mciteBstWouldAddEndPuncttrue
\mciteSetBstMidEndSepPunct{\mcitedefaultmidpunct}
{\mcitedefaultendpunct}{\mcitedefaultseppunct}\relax
\EndOfBibitem
\bibitem[Lin \latin{et~al.}(2016)Lin, Li, Ren, Koh, Kang, Peng, Xu, and
  Liu]{Lin2016}
Lin,~L.; Li,~J.; Ren,~H.; Koh,~A.~L.; Kang,~N.; Peng,~H. \latin{et~al.}
  \emph{ACS Nano} \textbf{2016}, \emph{10}, 2922--2929\relax
\mciteBstWouldAddEndPuncttrue
\mciteSetBstMidEndSepPunct{\mcitedefaultmidpunct}
{\mcitedefaultendpunct}{\mcitedefaultseppunct}\relax
\EndOfBibitem
\bibitem[Wu \latin{et~al.}(2015)Wu, Zhang, Yuan, Xue, Lu, Liu, Wang, Wang,
  Ding, Yu, Xie, and Jiang]{Wu2015}
Wu,~T.; Zhang,~X.; Yuan,~Q.; Xue,~J.; Lu,~G.; Liu,~Z. \latin{et~al.}
  \emph{Nat. Mater.} \textbf{2015}, \emph{15}, 43--47\relax
\mciteBstWouldAddEndPuncttrue
\mciteSetBstMidEndSepPunct{\mcitedefaultmidpunct}
{\mcitedefaultendpunct}{\mcitedefaultseppunct}\relax
\EndOfBibitem
\bibitem[Cummings \latin{et~al.}(2014)Cummings, Duong, Nguyen, Van~Tuan,
  Kotakoski, Barrios~Vargas, Lee, and Roche]{Cummings2014}
Cummings,~A.~W.; Duong,~D.~L.; Nguyen,~V.~L.; Van~Tuan,~D.; Kotakoski,~J.;
  Barrios~Vargas,~J.~E. \latin{et~al.}  \emph{Adv. Mater.} \textbf{2014},
  \emph{26}, 5079--5094\relax
\mciteBstWouldAddEndPuncttrue
\mciteSetBstMidEndSepPunct{\mcitedefaultmidpunct}
{\mcitedefaultendpunct}{\mcitedefaultseppunct}\relax
\EndOfBibitem
\bibitem[Mesaros \latin{et~al.}(2010)Mesaros, Papanikolaou, Flipse, Sadri, and
  Zaanen]{Mesaros2010}
Mesaros,~A.; Papanikolaou,~S.; Flipse,~C. F.~J.; Sadri,~D.; Zaanen,~J.
  \emph{Phys. Rev. B} \textbf{2010}, \emph{82}, 205119\relax
\mciteBstWouldAddEndPuncttrue
\mciteSetBstMidEndSepPunct{\mcitedefaultmidpunct}
{\mcitedefaultendpunct}{\mcitedefaultseppunct}\relax
\EndOfBibitem
\bibitem[Huang \latin{et~al.}(2011)Huang, Ruiz-Vargas, van~der Zande, Whitney,
  Levendorf, Kevek, Garg, Alden, Hustedt, Zhu, Park, McEuen, and
  Muller]{Huang2011}
Huang,~P.~Y.; Ruiz-Vargas,~C.~S.; van~der Zande,~A.~M.; Whitney,~W.~S.;
  Levendorf,~M.~P.; Kevek,~J.~W. \latin{et~al.}  \emph{Nature} \textbf{2011},
  \emph{469}, 389--393\relax
\mciteBstWouldAddEndPuncttrue
\mciteSetBstMidEndSepPunct{\mcitedefaultmidpunct}
{\mcitedefaultendpunct}{\mcitedefaultseppunct}\relax
\EndOfBibitem
\bibitem[Kim \latin{et~al.}(2011)Kim, Lee, Regan, Kisielowski, Crommie, and
  Zettl]{Kim2011}
Kim,~K.; Lee,~Z.; Regan,~W.; Kisielowski,~C.; Crommie,~M.~F.; Zettl,~A.
  \emph{ACS Nano} \textbf{2011}, \emph{5}, 2142--2146\relax
\mciteBstWouldAddEndPuncttrue
\mciteSetBstMidEndSepPunct{\mcitedefaultmidpunct}
{\mcitedefaultendpunct}{\mcitedefaultseppunct}\relax
\EndOfBibitem
\bibitem[Yazyev and Louie(2010)Yazyev, and Louie]{Yazyev2010}
Yazyev,~O.~V.; Louie,~S.~G. \emph{Nature Mater.} \textbf{2010}, \emph{9},
  806--809\relax
\mciteBstWouldAddEndPuncttrue
\mciteSetBstMidEndSepPunct{\mcitedefaultmidpunct}
{\mcitedefaultendpunct}{\mcitedefaultseppunct}\relax
\EndOfBibitem
\bibitem[Yu \latin{et~al.}(2011)Yu, Jauregui, Wu, Colby, Tian, Su, Cao, Liu,
  Pandey, Wei, Chung, Peng, Guisinger, Stach, Bao, Pei, and Chen]{Yu2011}
Yu,~Q.; Jauregui,~L.~A.; Wu,~W.; Colby,~R.; Tian,~J.; Su,~Z. \latin{et~al.}
  \emph{Nat. Mater.} \textbf{2011}, \emph{10}, 443--449\relax
\mciteBstWouldAddEndPuncttrue
\mciteSetBstMidEndSepPunct{\mcitedefaultmidpunct}
{\mcitedefaultendpunct}{\mcitedefaultseppunct}\relax
\EndOfBibitem
\bibitem[Tsen \latin{et~al.}(2012)Tsen, Brown, Levendorf, Ghahari, Huang,
  Havener, Ruiz-Vargas, Muller, Kim, and Park]{Tsen2012}
Tsen,~A.~W.; Brown,~L.; Levendorf,~M.~P.; Ghahari,~F.; Huang,~P.~Y.;
  Havener,~R.~W. \latin{et~al.}  \emph{Science} \textbf{2012}, \emph{336},
  1143--1146\relax
\mciteBstWouldAddEndPuncttrue
\mciteSetBstMidEndSepPunct{\mcitedefaultmidpunct}
{\mcitedefaultendpunct}{\mcitedefaultseppunct}\relax
\EndOfBibitem
\bibitem[Koepke \latin{et~al.}(2013)Koepke, Wood, Estrada, Ong, He, Pop, and
  Lyding]{Koepke2013}
Koepke,~J.~C.; Wood,~J.~D.; Estrada,~D.; Ong,~Z.-Y.; He,~K.~T.; Pop,~E.
  \latin{et~al.}  \emph{ACS Nano} \textbf{2013}, \emph{7}, 75--86\relax
\mciteBstWouldAddEndPuncttrue
\mciteSetBstMidEndSepPunct{\mcitedefaultmidpunct}
{\mcitedefaultendpunct}{\mcitedefaultseppunct}\relax
\EndOfBibitem
\bibitem[Gargiulo and Yazyev(2014)Gargiulo, and Yazyev]{Gargiulo2014nanolett}
Gargiulo,~F.; Yazyev,~O.~V. \emph{Nano Lett.} \textbf{2014}, \emph{14},
  250--254\relax
\mciteBstWouldAddEndPuncttrue
\mciteSetBstMidEndSepPunct{\mcitedefaultmidpunct}
{\mcitedefaultendpunct}{\mcitedefaultseppunct}\relax
\EndOfBibitem
\bibitem[Hung~Nguyen \latin{et~al.}(2016)Hung~Nguyen, Hoang, Dollfus, and
  Charlier]{Nguyen2016}
Hung~Nguyen,~V.; Hoang,~T.~X.; Dollfus,~P.; Charlier,~J.-C. \emph{Nanoscale}
  \textbf{2016}, \emph{8}, 11658--11673\relax
\mciteBstWouldAddEndPuncttrue
\mciteSetBstMidEndSepPunct{\mcitedefaultmidpunct}
{\mcitedefaultendpunct}{\mcitedefaultseppunct}\relax
\EndOfBibitem
\bibitem[Salehi-Khojin \latin{et~al.}(2012)Salehi-Khojin, Estrada, Lin, Bae,
  Xiong, Pop, and Masel]{Salehi2012}
Salehi-Khojin,~A.; Estrada,~D.; Lin,~K.~Y.; Bae,~M.-H.; Xiong,~F.; Pop,~E.
  \latin{et~al.}  \emph{Adv. Mater.} \textbf{2012}, \emph{24}, 53--57\relax
\mciteBstWouldAddEndPuncttrue
\mciteSetBstMidEndSepPunct{\mcitedefaultmidpunct}
{\mcitedefaultendpunct}{\mcitedefaultseppunct}\relax
\EndOfBibitem
\bibitem[Yasaei \latin{et~al.}(2014)Yasaei, Kumar, Hantehzadeh, Kayyalha,
  Baskin, Repnin, Wang, Klie, Chen, Kr\'{a}l, and Salehi-Khojin]{Yasaei2014}
Yasaei,~P.; Kumar,~B.; Hantehzadeh,~R.; Kayyalha,~M.; Baskin,~A.; Repnin,~N.
  \latin{et~al.}  \emph{Nat. Commun.} \textbf{2014}, \emph{5}, 4911\relax
\mciteBstWouldAddEndPuncttrue
\mciteSetBstMidEndSepPunct{\mcitedefaultmidpunct}
{\mcitedefaultendpunct}{\mcitedefaultseppunct}\relax
\EndOfBibitem
\bibitem[Seifert \latin{et~al.}(2015)Seifert, Vargas, Bobinger, Sachsenhauser,
  Cummings, Roche, and Garrido]{Seifert2015}
Seifert,~M.; Vargas,~J. E.~B.; Bobinger,~M.; Sachsenhauser,~M.;
  Cummings,~A.~W.; Roche,~S. \latin{et~al.}  \emph{2D Materials} \textbf{2015},
  \emph{2}, 024008\relax
\mciteBstWouldAddEndPuncttrue
\mciteSetBstMidEndSepPunct{\mcitedefaultmidpunct}
{\mcitedefaultendpunct}{\mcitedefaultseppunct}\relax
\EndOfBibitem
\bibitem[Boukhvalov and Katsnelson(2008)Boukhvalov, and
  Katsnelson]{Boukhvalov2008}
Boukhvalov,~D.~W.; Katsnelson,~M.~I. \emph{Nano Lett.} \textbf{2008}, \emph{8},
  4373--4379\relax
\mciteBstWouldAddEndPuncttrue
\mciteSetBstMidEndSepPunct{\mcitedefaultmidpunct}
{\mcitedefaultendpunct}{\mcitedefaultseppunct}\relax
\EndOfBibitem
\bibitem[OuYang \latin{et~al.}(2008)OuYang, Huang, Li, Xiao, Wang, and
  Xu]{OuYang2008}
OuYang,~F.; Huang,~B.; Li,~Z.; Xiao,~J.; Wang,~H.; Xu,~H. \emph{J. Phys. Chem.
  C} \textbf{2008}, \emph{112}, 12003--12007\relax
\mciteBstWouldAddEndPuncttrue
\mciteSetBstMidEndSepPunct{\mcitedefaultmidpunct}
{\mcitedefaultendpunct}{\mcitedefaultseppunct}\relax
\EndOfBibitem
\bibitem[Soler \latin{et~al.}(2002)Soler, Artacho, Gale, Garcia, Junquera,
  Ordejon, and Sanchez-Portal]{Soler2002}
Soler,~J.~M.; Artacho,~E.; Gale,~J.~D.; Garcia,~A.; Junquera,~J.; Ordejon,~P.
  \latin{et~al.}  \emph{J. Phys.: Condens. Matter} \textbf{2002}, \emph{14},
  2745\relax
\mciteBstWouldAddEndPuncttrue
\mciteSetBstMidEndSepPunct{\mcitedefaultmidpunct}
{\mcitedefaultendpunct}{\mcitedefaultseppunct}\relax
\EndOfBibitem
\bibitem[Perdew \latin{et~al.}(1996)Perdew, Burke, and Ernzerhof]{Perdew1996}
Perdew,~J.; Burke,~K.; Ernzerhof,~M. \emph{Phys. Rev. Lett.} \textbf{1996},
  \emph{77}, 3865--3868\relax
\mciteBstWouldAddEndPuncttrue
\mciteSetBstMidEndSepPunct{\mcitedefaultmidpunct}
{\mcitedefaultendpunct}{\mcitedefaultseppunct}\relax
\EndOfBibitem
\bibitem[Gargiulo \latin{et~al.}(2014)Gargiulo, Aut\`es, Virk, Barthel,
  R\"osner, Toller, Wehling, and Yazyev]{Gargiulo2014}
Gargiulo,~F.; Aut\`es,~G.; Virk,~N.; Barthel,~S.; R\"osner,~M.; Toller,~L.
  R.~M. \latin{et~al.}  \emph{Phys. Rev. Lett.} \textbf{2014}, \emph{113},
  246601\relax
\mciteBstWouldAddEndPuncttrue
\mciteSetBstMidEndSepPunct{\mcitedefaultmidpunct}
{\mcitedefaultendpunct}{\mcitedefaultseppunct}\relax
\EndOfBibitem
\bibitem[Duplock \latin{et~al.}(2004)Duplock, Scheffler, and
  Lindan]{Duplock2004}
Duplock,~E.~J.; Scheffler,~M.; Lindan,~P. J.~D. \emph{Phys. Rev. Lett.}
  \textbf{2004}, \emph{92}, 225502\relax
\mciteBstWouldAddEndPuncttrue
\mciteSetBstMidEndSepPunct{\mcitedefaultmidpunct}
{\mcitedefaultendpunct}{\mcitedefaultseppunct}\relax
\EndOfBibitem
\bibitem[Brito \latin{et~al.}(2011)Brito, Kagimura, and Miwa]{Brito2011}
Brito,~W.~H.; Kagimura,~R.; Miwa,~R.~H. \emph{Appl. Phys. Lett.} \textbf{2011},
  \emph{98}\relax
\mciteBstWouldAddEndPuncttrue
\mciteSetBstMidEndSepPunct{\mcitedefaultmidpunct}
{\mcitedefaultendpunct}{\mcitedefaultseppunct}\relax
\EndOfBibitem
\bibitem[Lieb(1989)]{Lieb1989}
Lieb,~E.~H. \emph{Phys. Rev. Lett.} \textbf{1989}, \emph{62}, 1201--1204\relax
\mciteBstWouldAddEndPuncttrue
\mciteSetBstMidEndSepPunct{\mcitedefaultmidpunct}
{\mcitedefaultendpunct}{\mcitedefaultseppunct}\relax
\EndOfBibitem
\bibitem[Inui \latin{et~al.}(1994)Inui, Trugman, and
  Abrahams]{PhysRevB.49.3190}
Inui,~M.; Trugman,~S.~A.; Abrahams,~E. \emph{Phys. Rev. B} \textbf{1994},
  \emph{49}, 3190--3196\relax
\mciteBstWouldAddEndPuncttrue
\mciteSetBstMidEndSepPunct{\mcitedefaultmidpunct}
{\mcitedefaultendpunct}{\mcitedefaultseppunct}\relax
\EndOfBibitem
\bibitem[Tuan \latin{et~al.}(2013)Tuan, Kotakoski, Louvet, Ortmann, Meyer, and
  Roche]{VanTuan2013}
Tuan,~D.~V.; Kotakoski,~J.; Louvet,~T.; Ortmann,~F.; Meyer,~J.~C.; Roche,~S.
  \emph{Nano Lett.} \textbf{2013}, \emph{13}, 1730--1735\relax
\mciteBstWouldAddEndPuncttrue
\mciteSetBstMidEndSepPunct{\mcitedefaultmidpunct}
{\mcitedefaultendpunct}{\mcitedefaultseppunct}\relax
\EndOfBibitem
\bibitem[Roche(1999)]{Roche1999}
Roche,~S. \emph{Phys. Rev. B} \textbf{1999}, \emph{59}, 2284--2291\relax
\mciteBstWouldAddEndPuncttrue
\mciteSetBstMidEndSepPunct{\mcitedefaultmidpunct}
{\mcitedefaultendpunct}{\mcitedefaultseppunct}\relax
\EndOfBibitem
\bibitem[Torres \latin{et~al.}(2014)Torres, Roche, and Charlier]{RocheBook}
Torres,~L. E. F.~F.; Roche,~S.; Charlier,~J.-C. \emph{Introduction to
  Graphene-Based Nanomaterials}; Cambridge University Press: Cambridge, UK,
  2014\relax
\mciteBstWouldAddEndPuncttrue
\mciteSetBstMidEndSepPunct{\mcitedefaultmidpunct}
{\mcitedefaultendpunct}{\mcitedefaultseppunct}\relax
\EndOfBibitem
\bibitem[Iitaka and Ebisuzaki(2004)Iitaka, and Ebisuzaki]{Iitaka2004}
Iitaka,~T.; Ebisuzaki,~T. \emph{Phys. Rev. E} \textbf{2004}, \emph{69},
  057701\relax
\mciteBstWouldAddEndPuncttrue
\mciteSetBstMidEndSepPunct{\mcitedefaultmidpunct}
{\mcitedefaultendpunct}{\mcitedefaultseppunct}\relax
\EndOfBibitem
\bibitem[Wei\ss{}e \latin{et~al.}(2006)Wei\ss{}e, Wellein, Alvermann, and
  Fehske]{Weisse2006}
Wei\ss{}e,~A.; Wellein,~G.; Alvermann,~A.; Fehske,~H. \emph{Rev. Mod. Phys.}
  \textbf{2006}, \emph{78}, 275--306\relax
\mciteBstWouldAddEndPuncttrue
\mciteSetBstMidEndSepPunct{\mcitedefaultmidpunct}
{\mcitedefaultendpunct}{\mcitedefaultseppunct}\relax
\EndOfBibitem
\bibitem[Zhang \latin{et~al.}(2014)Zhang, Lee, Gong, Colombo, and
  Cho]{Zhang2014a}
Zhang,~H.; Lee,~G.; Gong,~C.; Colombo,~L.; Cho,~K. \emph{J. Phys. Chem. C}
  \textbf{2014}, \emph{118}, 2338--2343\relax
\mciteBstWouldAddEndPuncttrue
\mciteSetBstMidEndSepPunct{\mcitedefaultmidpunct}
{\mcitedefaultendpunct}{\mcitedefaultseppunct}\relax
\EndOfBibitem
\bibitem[Santos \latin{et~al.}(2012)Santos, Ayuela, and
  Sánchez-Portal]{Santos2012}
Santos,~E. J.~G.; Ayuela,~A.; Sánchez-Portal,~D. \emph{New J. Phys.}
  \textbf{2012}, \emph{14}, 043022\relax
\mciteBstWouldAddEndPuncttrue
\mciteSetBstMidEndSepPunct{\mcitedefaultmidpunct}
{\mcitedefaultendpunct}{\mcitedefaultseppunct}\relax
\EndOfBibitem
\bibitem[Kochan \latin{et~al.}(2014)Kochan, Gmitra, and Fabian]{Kochan2014}
Kochan,~D.; Gmitra,~M.; Fabian,~J. \emph{Phys. Rev. Lett.} \textbf{2014},
  \emph{112}, 116602\relax
\mciteBstWouldAddEndPuncttrue
\mciteSetBstMidEndSepPunct{\mcitedefaultmidpunct}
{\mcitedefaultendpunct}{\mcitedefaultseppunct}\relax
\EndOfBibitem
\bibitem[Soriano \latin{et~al.}(2011)Soriano, Leconte, Ordej\'on, Charlier,
  Palacios, and Roche]{Soriano2011}
Soriano,~D.; Leconte,~N.; Ordej\'on,~P.; Charlier,~J.-C.; Palacios,~J.-J.;
  Roche,~S. \emph{Phys. Rev. Lett.} \textbf{2011}, \emph{107}, 016602\relax
\mciteBstWouldAddEndPuncttrue
\mciteSetBstMidEndSepPunct{\mcitedefaultmidpunct}
{\mcitedefaultendpunct}{\mcitedefaultseppunct}\relax
\EndOfBibitem
\end{mcitethebibliography}

\end{document}